\documentclass{epl}

\newcommand{\ie}{{\em i.e. }}

\title{Diphasic non-local model for granular surface flows}

\shorttitle{Diphasic non-local model for granular surface flows}

\author{D. Bonamy\inst{1,2} \and P. Mills\inst{3}}

\shortauthor{D. Bonamy \and P. Mills}

\institute{
  \inst{1} Service de Physique de l'Etat Condens\'e - CEA Saclay, 91191 Gif-sur-Yvette 
Cedex, France\\
  \inst{2} Laboratoire des Verres - Universit{\'e} Montpellier 2 - F-34095 Montpellier Cedex
5, France\\
  \inst{3} Laboratoire de Physique des Mat{\'e}riaux Divis{\'e}s et des Interfaces - Universit{\'e}
Marne la Vall{\'e}e, 77454 Marne la Vall{\'e}e Cedex 2, France\\
}

\begin{document}

\maketitle

\begin{abstract}
Considering recent results revealing the existence of multi-scale rigid clusters of grains embedded in granular surface flows, {\em i.e.} flows down an erodible bed, we describe here the surface flows rheology through a non-local constitutive law. The predictions of the resulting model are compared quantitatively to experimental results: The model succeeds to account for the counter-intuitive shape of the velocity profile observed in experiments, {\em i.e.} a velocity profile decreasing exponentially with depth in the static phase and remaining linear in the flowing layer with a velocity gradient independent of both the flowing layer thickness, the angle between the flow and the horizontal, and the coefficient of restitution of the grains. Moreover, the scalings observed in rotating drums are recovered, at least for small rotating speed. 
\end{abstract}

\section{Introduction}

Granular media share properties with both usual liquids and solids: They can sustain an inclined free surface without flowing but, when the angle exceeds a critical value identified since Coulomb~\cite{Coulomb76} with some effective macroscopic friction angle $\Phi_M$, an avalanche occurs. The motion has the peculiarity of being a surface flow. Two phases can be observed: A "solid" phase experiencing a creep motion where the averaged streamwise velocity decreases exponentially with the depth~\cite{Komatsu01,Bonamy02}, and a flowing phase exhibiting a linear velocity profile with a velocity gradient independent of the flowing layer thickness, the angle between the mean flow and the horizontal, and the coefficient of restitution of the beads~\cite{Rajchenbach00,Bonamy02}. Such profiles cannot be described using any conventional local and univocal constitutive laws of Continuum Mechanics for two main reasons: (i) The velocity gradient is found to be constant in the cascading layer whereas momentum balance implies that the shear stress increases linearly with depth and (ii) the velocity profile in the flowing layer down an erodible bed differs significantly from the Bagnold-like velocity profile observed in dense flows down rough inclines~\cite{RoughIncline}. Several models have been recently proposed to describe dense granular flows~\cite{Pouliquen02}, but, to our knowledge, none of them succeeds to account for the velocity profile measured experimentally in surface flows.

A very recent experiment~\cite{Bonamy02b} has provided evidence of multi-scale rigid clusters embedded in the flow that may be responsible for non-local interactions in surface flows. Consequently, following the work of Mills {\em et al.}~\cite{Mills99} originally devoted to dense flow down a rough incline, we propose here to describe the surface flow rheology through a non-local constitutive law. The resulting model is presented and discussed in the first section. In the second section, the model is applied to the description of {\em ideal} 2D flows at the surface of a packing of infinite depth. Among others, it accounts for the shape of the velocity profile observed experimentally. Finally, in the last section, we focus more specifically on steady surface flows in a {\em slowly} rotating {\em large} drum of {\em finite} thickness. The resulting model describes the experimental observations successfully.

\section{General formulation}

\begin{figure}
\onefigure[width=0.6\textwidth]{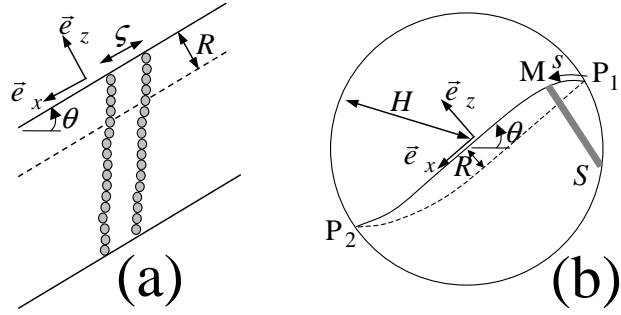}
\caption{(a): Notation and sketch for the 2D surface flows The quantities $R$ and $\theta$
refer respectively to the flowing layer thickness and the mean angle of the flow.
the transient solid objects are assumed to be separated by a mean distance $\zeta$ constant in the whole packing. The frame $(\vect{e}_x,\vect{e}_z)$ is chosen tangent to the free surface. (b) Rotating drum geometry. The radius of the drum and the solid rotation velocity are called respectively $H$ and $\Omega$. The origin of $(\vect{e}_x,\vect{e}_z)$ is now chosen to coincide with the center of the drum. For $x=0$, $\partial_x \nu$ and $\partial_x v_x$ can be assumed to vanish, but $\partial_x v_z$ do not. The points $\mathrm{P}_1$ $\mathrm{P}_2$ denote respectively the upper and the lower boundary of the flowing layer. A point M of the free surface is defined through its curvilinear coordinate $s$ whose origin is set at $\mathrm{P}_1$. The surface $S$ is normal to the free surface at M and bounded by the free surface and the drum boundary.}
\label{fig1}
\end{figure}

Let us consider surface flows of monodisperse spheres of diameter $d$ and of density $\rho_0$ in the gravity field $\vect{g}$. Assuming that granular media deal with continuum models, the velocity and stress field should verify the momentum equation:
$\rho_0 \nu \{\partial_t \vect{v}+(\vect{v}.\vect{\nabla})\vect{v}\}=\rho_0
\nu \vect{g}+\vect{\nabla}.\tens{\Sigma}$ where $\vect{v}$ is the velocity of the fluid particle, $\nu$ the volume fraction and $\tens{\Sigma}$ the stress tensor. In the following, variables are non-dimensionalized by length $d$, time $\sqrt{d/g}$, and stress $\rho_0 g d$. Moreover, only steady, homogeneous 2D surface flows are considered:
In the frame $(\vect{e}_x,\vect{e}_z)$ tangent to the free surface (Fig. \ref{fig1}a),
$\vect{v}(x,z,t)=v(z) \vect{e}_x$, $\nu(x,z,t)=\nu(z)$ and the projection of the conservation equations gives:

\begin{equation}
(a)\quad \frac{\upd \Sigma_{zz}}{\upd z}=\nu \cos \theta, \quad
(b)\quad \frac{\upd \Sigma_{xz}}{\upd z}=- \nu \sin \theta
\label{equ1}
\end{equation}

A constitutive law should now be provided to relate these stress profiles to "observables" profiles such as velocity and volume fraction profiles. Correlated rigid clusters of grains embedded in surface flows have been experimentally evidenced~\cite{Bonamy02b}. Their size has the peculiarity to be power-law distributed from the grain diameter to the flowing layer thickness. Consequently no characteristic correlation length scale -~and thus no local stress/strain constitutive relationship~- can be defined. A non-local rheological law is thus proposed below.

Following Mills {\em et al.}~\cite{Mills99}, we express the stress components $\Sigma_{ij}$ as the sum of a motionless part $\Sigma_{ij}^{(s)}$ and an hydrodynamic part $\Sigma_{ij}^{(d)}$. The motionless components are assumed to keep the relationship they got just before static equilibrium failure and the hydrodynamic parts are evaluated by considering the flow as transient solid chains embedded in a viscous fluid. This leads to (see~\cite{Mills99} for justifications and details):

\begin{equation}
\begin{array}{ll}

(a) \quad \Sigma_{zz}=\Sigma_{zz}^{(s)}+\Sigma_{zz}^{(d)} \quad & (b) \quad \Sigma_{xz}=\Sigma_{xz}^{(s)}+\Sigma_{xz}^{(d)}\\

(c) \quad \frac{\upd \Sigma^{(s)}_{zz}}{\upd z}=\nu_0 \cos \theta \quad & (d) \quad \Sigma^{(s)}_{xz} = -\tan\Phi \Sigma^{(s)}_{zz}\\

(e) \quad \Sigma^{(d)}_{zz}= \alpha_0 \frac{\tan\Phi}{\zeta} \int_{z}^{0}\sigma^v(y)dy &
(f) \quad \Sigma^{(d)}_{xz}= \sigma^v + \frac{1}{\zeta} \int_{z}^{0}\sigma^v(y)dy

\end{array}
\label{equ2}
\end{equation}

\noindent where $\alpha_0$ is a constant coefficient related to the mean transient chains orientations~\cite{Mills99}, $\zeta$ the correlation length of the chain network in a shear plane $z=C^{te}$, $\Phi$ the internal friction angle of the material, $\sigma^{v}=\eta \upd v_x/ \upd z$ the shear stress in the viscous fluid and $\nu_0$ the volume fraction of the immobile stack before failure that ranges from the random loose packed limit $\nu^*_l \simeq 0.55$ to the random close packed limit $\nu^*_c \simeq 0.64$. Since the influence of the initial volume fraction $\nu_0$ on the rheology is not investigated in this paper, we set $\nu_0=\nu^*_c$ in the following.

The internal friction angle $\Phi$ is the sum of the particle/particle friction angle $\Phi_m$ and of an angle $\beta$ reflecting the entanglement of the grains~\cite{Rowe62}. According to the Reynolds dilantancy principle~\cite{Reynolds85} $\beta$ vanishes when the volume fraction is smaller than the random loose packed value $\nu^*_l$ and increases with $\nu$ when $\nu>\nu^*_l$ to be a maximum when $\nu=\nu^*_c$. Consequently, $\Phi(\nu)= \Phi_m+(\Phi_M-\Phi_m)f((\nu-\nu^*_l)/(\nu^*_c-\nu^*_l))$ where the maximum value $\Phi_M$ of the internal friction angle can be identified with the angle of repose and $f(x)$ is a dimensionless function increasing with repsect to $x$ so that $f(0)=0$ and $f(1)=1$. For sake of simplicity, $f$ is identified with an Heaviside function to allow analytical solutions, \ie:

\begin{equation}
\mathrm{for} \quad \nu \leq \nu^*_l \quad \Phi=\Phi_m, \quad
\mathrm{for} \quad \nu > \nu^*_l \quad \Phi=\Phi_M
\label{equ3}
\end{equation}

Expressions for the viscosity are usually obtained from the Enskog equation~\cite{Jenkins83}. It reads $\eta\propto g(\nu) \sqrt{T}$ where $g(\nu)$ is the pair correlation function expected to diverge as $(\nu^*_c-\nu)^{-1}$ in the high density limit~\cite{Speedy99} and $\sqrt{T}$ is the granular temperature identified with the rms velocity fluctuations. For dilute flows, the temperature profile can be determined by a {\em local} balance between the viscous heating and dissipation through inelastic binary collisions~\cite{Jenkins83}. But in dense flows, the kinetic energy of a collision is dissipated within the entire packing through multiple collisions and the concept of a {\em local} energy balance becomes irrelevant~\cite{Rajchenbach00,Bonamy02,Andreotti01}. Moreover, the mean free path is very small and the experimental noise is expected to be sufficient to drive collisions~\cite{deGennes95}. For all these reasons, we choose here to define an effective temperature
$T_{eff}$ {\em constant in the whole packing}, resulting from the global energy balance within the whole system. The viscosity can then be written as:

\begin{equation}
\sigma^\nu = \eta_0 g(\nu) \frac{\upd v_x}{\upd z} \quad \mathrm{where} \quad g(\nu) = \frac{1}{\nu^*_c-\nu}
\label{equ4}
\end{equation}
 
Finally, from Eqs.~\ref{equ1} and \ref{equ2}, one can relate $\nu$ to $\sigma^v$ via:

\begin{equation}
\nu=\nu^*_c-\frac{\alpha_0 \tan\Phi}{\zeta \cos\theta} \sigma^v
\label{equ5}
\end{equation}

\section{Canonical case: Flows down to the surface of an infinite packing}

Let us consider now the granular flow at the surface of a packing of infinite depth and infinite width. From Eqs.~\ref{equ1} and \ref{equ2}, the viscous stress $\sigma^v$ should obey:

\begin{equation}
\frac{\upd \sigma^v }{\upd z}-\frac{1}{\lambda(\theta,\Phi)}\sigma^v = - \nu^*_c F(\theta,\Phi)
\label{equ6}
\end{equation}

\noindent where the functions $\lambda$ and $F$ are given respectively by: $F(\theta_1,\theta_2)=\cos\theta_1(\tan\theta_1 - \tan\theta_2)$ and $\lambda(\theta_1,\theta_2)=\zeta/(1+\alpha_0\tan\theta_1\tan\theta_2)$. For sake of simplicity, $\theta$ variations are assumed to be small around $\Phi_{M}$ and the different trigonometric functions are expanded in the relevant order of $\theta-\Phi_{M}$. The expressions of $F$ and $\lambda$ become: $F(\theta_1,\theta_2)\simeq (\theta_1 - \theta_2)/\cos\Phi_M$ and $\lambda(\theta_1,\theta_2) \simeq \zeta/(1+\alpha_0\tan^2\Phi_M) \simeq C^{te}$.

For a surface flow, $\sigma^v(z)$ is expected to vanish for infinite depth. From Eq.~\ref{equ5}, Eq.~\ref{equ3} and Eq.~\ref{equ6}, one deduces sucessively that $\nu \rightarrow \nu^*_c$, $\Phi \rightarrow \Phi_M$ and $F(\theta,\Phi_M) \rightarrow 0$ when $z \rightarrow -\infty$. Consequently, permanent homogeneous surface flows are available in infinite packing only for $\theta=\Phi_M$. When $z$ is gradually increased from $-\infty$, first $\Phi=\Phi_M$ (Eq.~\ref{equ3}),  $\sigma^v$ increases (Eq.~\ref{equ6}) and $\nu$ decreases (Eq.~\ref{equ5}). Then, at a given depth identified with the cascading layer/erodible bed interface, $\nu=\nu^*_l$, which switches the value of $\Phi=\Phi_M$ to $\Phi=\Phi_m$ (Eq.~\ref{equ3}). Consequently, Eq.~\ref{equ6} can be written:

\begin{equation}
\begin{array}{ll}
\mathrm{for} \quad z\in [-\infty,-R] \quad & \frac{\upd \sigma^v }{\upd z}-\frac{1}{\lambda}\sigma^v = 0\\
\mathrm{for} \quad z\in [-R,0] \quad & \frac{\upd \sigma^v }{\upd z}-\frac{1}{\lambda}\sigma^v = - \nu^*_c F(\Phi_M,\Phi_m)
\end{array}
\label{equ7}
\end{equation}

\noindent where $R$ is the flowing layer thickness. Let us note that, in this approach, the presence of the two phases, the "static" phase and the flowing layer, comes from the variation of $\Phi$ with respect to $\nu$. Then, as $\Sigma_{ij}^{(s)}$, $\Sigma_{ij}^{(d)}$ and consequently $\sigma^v$ (Eq.~\ref{equ2}) should vanish at the free surface, and as $\Sigma_{ij}^{(d)}$, and consequently $\sigma^v$ (Eq.~\ref{equ2}) should vary continuously at the interface~\cite{matchingCondition}, one gets:

\begin{equation}
\begin{array}{ll}
\mathrm{for}\: z\in [-\infty,-R] \: & \sigma^v(z)=\nu^*_c \lambda F(\Phi_M,\Phi_m)(e^{R/\lambda}-1)e^{z/\lambda}\\
\mathrm{for}\: z\in [-R,0] \: & \sigma^v(z) = \nu^*_c \lambda F(\Phi_M,\Phi_m)(1-e^{z/\lambda})
\end{array}
\label{equ8}
\end{equation}

\noindent In the following, one assumes $\lambda \ll R$ and does not consider what happens in the vicinity of the free surface $-\lambda \leq z \leq 0$. The $\sigma^\nu$ profile is then given by:

\begin{equation}
\begin{array}{ll}
\mathrm{for} \quad z\in [-\infty,-R] \quad & \sigma^v(z) \simeq \nu^*_c \lambda F(\Phi_M,\Phi_m) e^{R/\lambda}e^{z/\lambda}\\
\mathrm{for} \quad z\in [-R,-\lambda] \quad & \sigma^v(z) \simeq \nu^*_c \lambda F(\Phi_M,\Phi_m)
\end{array}
\label{equ9}
\end{equation}

\noindent From Eqs.~\ref{equ5} and \ref{equ9} the volume fraction profile can be deduced:

\begin{equation}
\begin{array}{ll}
\mathrm{for} \quad z\in [-\infty,-R] \quad & \nu(z) = \nu^*_c\left(1-(1-\frac{\lambda}{\zeta})\frac{F(\Phi_M,\Phi_m)}{\sin\Phi_M}e^{R/\lambda}e^{z/\lambda}\right)\\
\mathrm{for} \quad z\in [-R,-\lambda] \quad & \nu(z) = \nu^*_c\left(1-(1-\frac{\lambda}{\zeta})\frac{F(\Phi_M,\Phi_m)}{\sin\Phi_M}\right)
\end{array}
\label{equ10}
\end{equation}

\noindent The volume fraction is thus predicted to be {\em different}, but {\em constant} in the two phases except in a narrow layer at the interface where the two values are connected through a matching exponential function of characteristic length $\lambda$. From Eqs.~\ref{equ4},\ref{equ9} and \ref{equ10} the velocity profile can be deduced:

\begin{equation}
\begin{array}{ll}
\mathrm{for} \quad z\in [-\infty,-R] \quad & v_x(z) = \frac{\lambda}{2} \dot{\gamma_0} e^{2R/\lambda}e^{2z/\lambda}\\
\mathrm{for} \quad z\in [-R,-\lambda] \quad & v_x(z) = \dot{\gamma_0}(z+R+\frac{\lambda}{2})
\end{array}
\label{equ11}
\end{equation}

\noindent where the velocity gradient $\dot{\gamma}_0$ is given by:

\begin{equation}
\dot{\gamma}_0=\frac{{\nu^*_c}^2}{\eta_0}\frac{F(\Phi_M,\Phi_m)^2}{\sin\Phi_M}\lambda(1-\frac{\lambda}{\zeta})
\label{equ12}
\end{equation}

\noindent which captures the main experimental observations, namely a velocity profile decreasing exponentially with depth in the static phase~\cite{Komatsu01,Bonamy02} and remaining linear in the flowing layer with a velocity gradient independent of $R$~\cite{Bonamy02,Rajchenbach00}.

Let us note finally that the use of a generic viscous law $\sigma^v=\eta_0(\upd v_x/\upd z)^\alpha/(\nu^*_c-\nu)^\beta$ modifies the shape of the velocity profile only through $\dot{\gamma}_0$ and the characteristic length of the exponential term that become respectively $\dot{\gamma}_0 \rightarrow \{{\nu^*_c}^{\beta+1} F(\Phi_M,\Phi_m)^{\beta+1}\lambda(1-\lambda/\zeta)^\beta/(\eta_0 \sin^\beta\Phi_M)\}^{1/\alpha}$ and $\lambda \rightarrow \alpha\lambda/(\beta+1)$. This indicates that the precise expression of the viscosity, and in particular the assumption of a constant temperature, has no influence on the shape of the velocity profile.

\section{Steady surface flows in a large rotating drum}

Let us now investigate permanent surface flows at the center of an half-filled rotating drum of radius $H$ (cf. Fig.~\ref{fig1}b). The origin of the frame $(\vec{e}_x,\vec{e}_z)$ is now chosen at the center of the drum with $\vec{e}_x$ (resp. $\vec{e}_z$) is tangent (resp. normal) to the free surface (see fig.~\ref{fig1}b). The rotating speed $\Omega$ is supposed to be small so that only first order terms in $\Omega$ are considered. The inertial stresses scale as $\Omega^2$ and can thus be neglected in Eq.~\ref{equ1}. In such geometry, the flow cannot be considered as homogeneous anymore. $\partial_x \nu$ and $\partial_x v_x$ vanish for $x=0$, but $\partial_x v_z$ do not: At the drum boundary, $\vec{v}$ is given by the solid rotation of the drum which leads to $\partial_x v_z (x=0;z=-H)=-\Omega$. More generally, one can write $v_z (\delta x;z)=-\Omega\delta x f(z)$ where $f$ is a dimensionless decreasing function of $z$ so that $f(-H)=1$ and $f(0)=0$. This introduces new terms $\Sigma^\nu_{xx}=\Sigma^\nu_{zz}=-\eta_\nu\nabla.\vec{v}=-\eta_\nu\partial_z v_z$ in the stress tensor representing the dissipation generated by dilation where $\eta_\nu$ is the volume viscosity and an unknown function of $\nu$. The conservation equation previously given by Eq.~\ref{equ1} becomes $\partial_x \Sigma^\nu_{xx}+\upd\Sigma_{xz}/\upd z=-\nu\sin\theta$ and $\upd\Sigma_{zz}/\upd z=\nu\cos\theta$. Then, $\sigma^\nu$ should obey:

\begin{equation}
\frac{\upd \sigma^v }{\upd z}-\frac{1}{\lambda}\sigma^v = - \nu^*_c F_\Omega(\theta,\Phi,\Omega,\nu,z)
\label{equ13}
\end{equation}

\noindent where $F_\Omega(\theta,\Phi,\Omega,\nu,z)=F(\theta,\Phi)+\eta_\nu(\nu)\Omega f'(z)/\nu^*_c$. As $\sigma^\nu(-H)$ vanishes, we deduce that $\nu(-H)=\nu^*_c$ (Eq.~\ref{equ5}), $\Phi(-H)=\Phi_M$ (Eq.~\ref{equ3}) and, assuming that $H\gg R$, $F_\Omega(\theta,\Phi_M,\Omega,\nu^*_c,-H)\simeq 0$ (Eq.~\ref{equ13}), \ie: 

\begin{equation}
\theta = \Phi_M + p\Omega \quad \mathrm{where} \quad p=\frac{\eta_\nu(\nu^*_c)}{\nu^*_c}\cos\Phi_M\mid f'(-H)\mid
\label{equ14}
\end{equation}

\noindent which reproduce the scaling relations $\theta(\Omega)$ observed experimentally when $\Omega$ is small enough to neglect inertial effects~\cite{Bonamy02,Orpe01}.

The forms of $f(z)$ and $\eta_\nu(\nu)$ are now needed to evaluate the velocity profile. For sake of simplicity, one chooses here the simplest form compatible with the boundary conditions $f(0)=0$ and $f(-H)=1$, \ie $f(z)=-z/H$. For the same reason, $\eta_B$ is also set constant in the static phase. Then, $\sigma^v$ obeys:

\begin{equation}
\begin{array}{ll}
\mathrm{for} \quad z\in [-H,-R] \quad & \frac{\upd \sigma^v }{\upd z}-\frac{1}{\lambda}\sigma^v = - \nu^*_c F(\theta,\Phi_M)+\eta_\nu\Omega/H=0\\
\mathrm{for} \quad z\in [-R,0] \quad & \frac{\upd \sigma^v }{\upd z}-\frac{1}{\lambda}\sigma^v = - \nu^*_c F(\theta,\Phi_m)+\eta_\nu\Omega/H= - \nu^*_c F(\Phi_M,\Phi_m)
\end{array}
\label{equ15}
\end{equation}

And the volume fraction profile (resp. the volume fraction profile) are given by Eqs.~\ref{equ8} and \ref{equ9} (resp. Eq.~\ref{equ10}). The viscous relationship now reads $\sigma^\nu=\eta_0 g(\nu) (\upd v_x/\upd z+\partial_x v_z)$ with $\partial_x v_z=\Omega z /H$ and, assuming that $H \gg R \gg \lambda$, the shape of the velocity profile is finally given by:

\begin{equation}
\begin{array}{ll}
\mathrm{for} \quad z\in [-H,-R] \quad & v_x(z) \simeq \frac{\lambda}{2} \dot{\gamma_0} e^{2R/\lambda}e^{2z/\lambda}-\frac{\Omega}{2 H}(z^2+H^2)\\
\mathrm{for} \quad z\in [-R,-\lambda] \quad & v_x(z) \simeq \dot{\gamma}_0\left(z+R+\frac{\lambda}{2}\right)-\frac{\Omega}{2H} \left(z^2+H^2\right)
\end{array}
\label{equ16}
\end{equation}

\noindent And consequently, for $\Omega \ll \dot{\gamma}_0$ as in experiments~\cite{Rajchenbach00,Bonamy02}, the solid rotation of the packing does not modify significantly the shape of the velocity profile. The relationship between the flowing layer thickness $R$ and $\Omega$ can now be deduced: For a steady flow, at each point M of the free surface, the flow rate $Q(\mathrm{M})$ through the surface $S(\mathrm{M})$ normal to the free surface at M and bounded by the free surface and the drum boundary (Fig.~\ref{fig1}b) should obey the mass conservation equation: $\partial_s Q= 0$ where $s$ is the curvilinear coordinate of the point M. As $Q$ vanishes at the upper and lower boundaries of the flowing layer (points $\mathrm{P}_1$ and $\mathrm{P}_2$ of Fig.~\ref{fig1}b), $Q$ vanishes everywhere, in particular at the center of the drum. One thus gets: $Q=\int_{-H}^{0} v_x(z)\upd z=0$, and, assuming $\lambda \ll R \ll H$:

\begin{equation}
R \simeq H\left( \frac{4\Omega}{3\dot{\gamma_0}}\right)^{1/2}
\label{equ17}
\end{equation}

\noindent which is in agreement with the experimental scaling reported in~\cite{Rajchenbach00,Bonamy02,Orpe01}.

\begin{figure}
\twoimages[width=6cm]{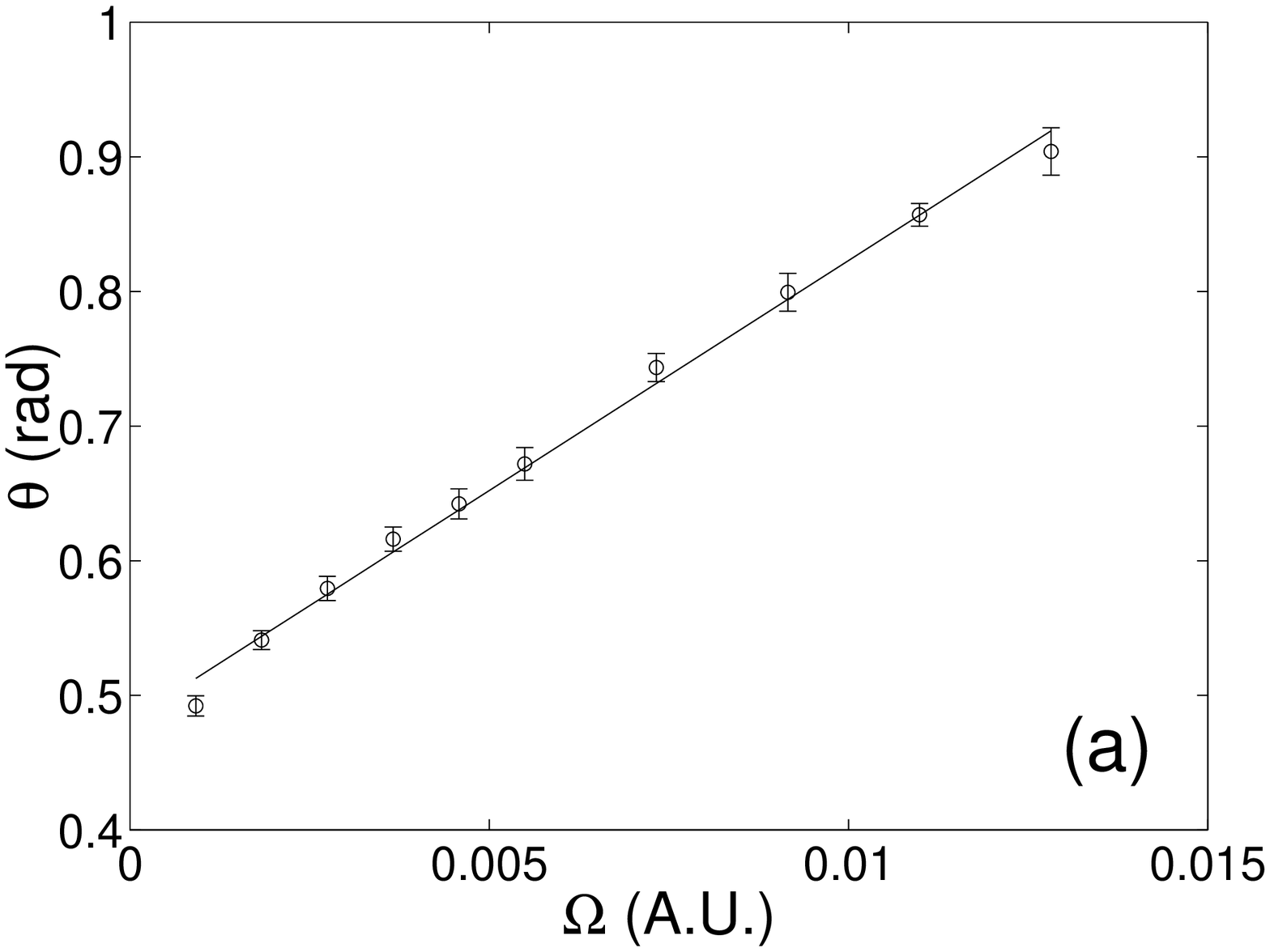}{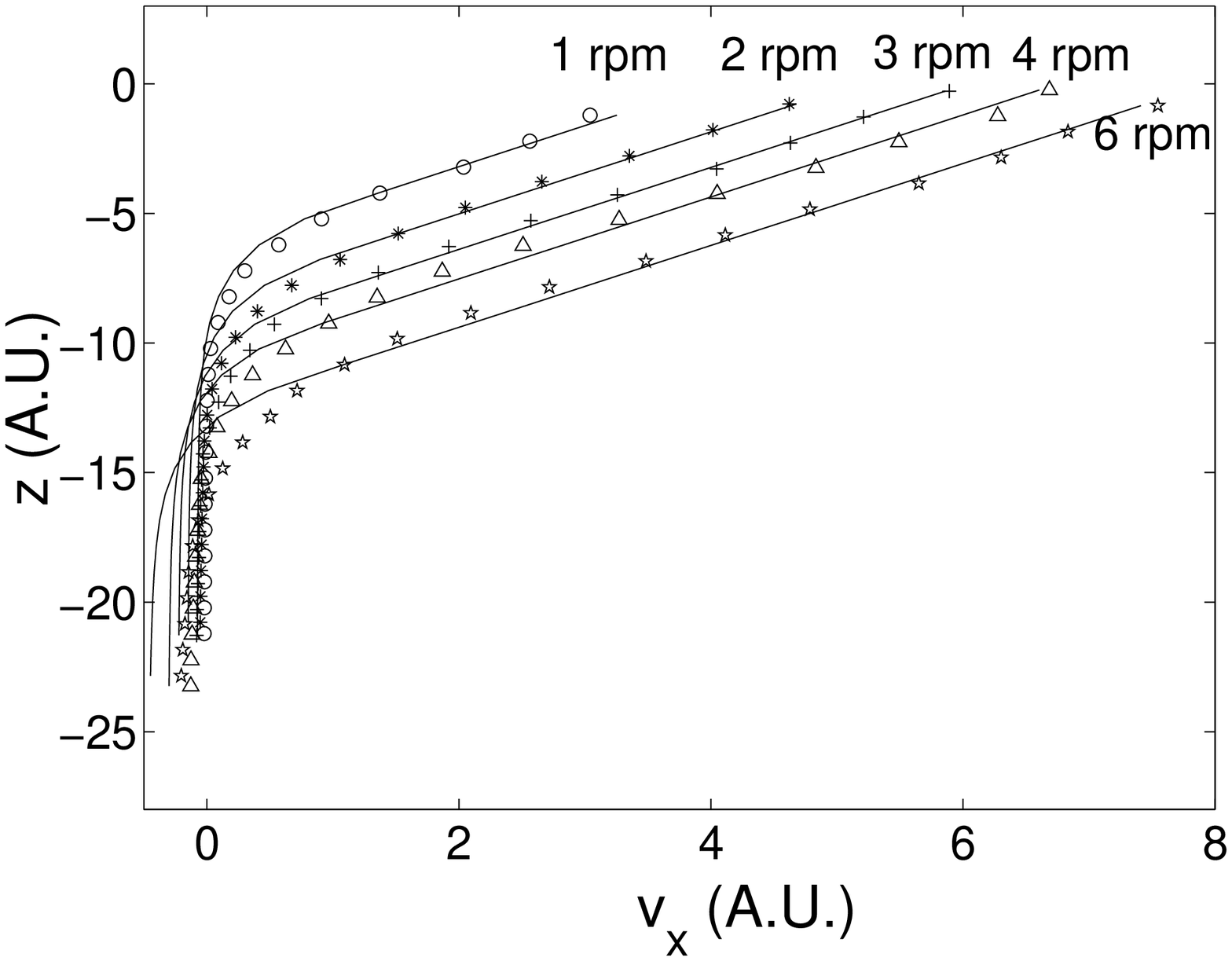}  
\caption{Comparison between theory and experiment.(a): Variation of the angle $\theta$ (in radian) of the mean flow {\em versus} the rotation speed $\Omega$ (non-dimensionalized by $\sqrt{g/d}$). Circles correspond to the experimental results of~\cite{Bonamy02} obtained in a rotating drum of diameter $H=45\un{cm}$ and of thickness $7\un{mm}$, half-filled with steel beads of diameter $d=3\un{mm}$. The critical angle $\Phi_M$ has been measured to be $\Phi_M=27.5^\circ$. The full line is the theoretical result given by Eq.~\ref{equ14} with a fit parameter $p=32.6$. (b): Velocity profile $v_x(z)$ in non-dimensionalized Units. Full lines corresponds to theoretical profiles given by Eq.~\ref{equ16}. Experimental data are fitted by changing the parameter $\dot{\gamma}_0$ for the entire set of profiles and by evaluating $R$ from Eq.~\ref{equ17} for each profile. The fitted correlation length and velocity gradient are found to be $\lambda=3.62$ and $\dot{\gamma}_0=0.63$ respectively.}
\label{fig2}
\end{figure}

In Fig.~\ref{fig2}, the model predictions are compared to the data  of~\cite{Bonamy02} measured experimentally in a rotating drum of diameter $H=45\un{cm}$ and of thickness $7\un{mm}$, half-filled with steel beads of diameter $d=3\un{mm}$ with a particle/particle friction angle $\Phi_m=10^\circ$. Moreover, the angle of repose has been measured in the experiment: $\Phi_M=27.5^\circ$. Figure~\ref{fig2}a presents the variation of $\theta$ with $\Omega$ measured experimentally in~\cite{Bonamy02} and calculated using the theoretical scaling (Eq.~\ref{equ14}). The fit parameter is found to be $p=32.6$. Figure~\ref{fig2}b presents the comparison between the velocity profile given by Eq.~\ref{equ16} and the ones measured experimentally in~\cite{Bonamy02}. The last ones are fitted by setting $H=75$, by evaluating $R$ using Eq.~\ref{equ17} for each profile, and by changing $\dot{\gamma_0}$ and $\lambda$ for the entire set of profiles. The agreement is remarkable. The value of the fit parameters are found to be $\dot{\gamma}_0=0.63$ and $\lambda=3.62$.

\section{Conclusion}

In this letter, we have proposed a non-local diphasic model somewhat derived from~\cite{Mills99} to describe surface flows in granular packing. This model accounts for the shape of the velocity profile experimentally observed, which no theories allowed for until now. Then, we have applied the model to the description of steady surface flows in a rotating drum. All the experimentally observed scaling were recovered, at least for rotating speed slow enough to let inertial effect be negligible. Our theoretical predictions agree quantitatively with the experimental results of~\cite{Bonamy02}.

A main limitation of the model comes from the assumption that the two lengths characterizing the granular mesostructure, \ie the length $\xi$ of the  transient chains (infinite here) and the average distance $\lambda$ between the chains in a shear plane, are fixed throughout the entire packing. Assuming connectivity transition in a finite size system, one can relate these two lengths to the volume fraction and the size of the media. This allows to extend the present non-local description to other geometries such as flows down rough inclines~\cite{RoughIncline} and shear Couette flows~\cite{Couette}. This will be explored in a forthcoming publication~\cite{Mills02}.

\acknowledgments

We acknowledge discussions with E. Bertin, F. Chevoir, F. Daviaud, O. Dauchot and L. Laurent. We also thank Dan Qvale for the careful reading of this manuscript.


\begin{thebibliography}{0}
\bibitem{matchingCondition}
Note that imposing the continuity of the velocity gradient at the interface induces a non continous shear stress $\Sigma_{xz}$. This comes from the choice of a non-continous friction angle $\Phi(\nu)$ that changes abruptly from when $\nu < \nu_m$ becomes $\nu > \nu^*_l$ at the interface.
\end{thebibliography}
\end{document}